\documentclass[sigconf]{acmart}
\pdfoutput=1
\usepackage{amsmath}
\usepackage{amssymb}
\usepackage{amsfonts}
\usepackage{algorithmic}
\usepackage{graphicx}
\usepackage{textcomp}
\def\BibTeX{{\rm B\kern-.05em{\sc i\kern-.025em b}\kern-.08em
    T\kern-.1667em\lower.7ex\hbox{E}\kern-.125emX}}
\usepackage{multirow}
\usepackage{tabularx}
\usepackage{booktabs}
\usepackage{amsmath}
\usepackage{amsthm}
\usepackage{color}
\usepackage{enumitem}

\usepackage{caption}
\usepackage{subcaption} 
\usepackage{listings}
\usepackage{pgfplots}
\usepackage{color}
\DeclareCaptionFormat{listing}{\colorbox{gray!50}{\parbox[l]{\columnwidth}{#1#2#3}}}

\newcommand{\foivos}[1]{\textbf{\color{blue}FT:{#1}}}

\begin{document}

\title{Learning to Encode and Classify Test Executions}
\author{Foivos Tsimpourlas} 
\affiliation{University of Edinburgh}
\email{F.Tsimpourlas@sms.ed.ac.uk}
\author{Ajitha Rajan}
\affiliation{University of Edinburgh}
\email{arajan@ed.ac.uk}
\author{Miltiadis Allamanis}
\affiliation{Microsoft Research}
\email{miallama@microsoft.com}

\begin{abstract}

The challenge of automatically determining the correctness of test executions is referred to as the \emph{test oracle problem} and is a
key remaining issue for automated testing.
The paper aims at solving the test oracle problem in a scalable and accurate way. 
To achieve this, we use supervised learning over test execution traces. We label a small fraction of the execution traces with their verdict of pass or fail. 
We use the labelled traces to train a neural network (NN) model to learn to distinguish runtime patterns for passing versus failing executions for a given program. 

We evaluate our approach using case studies from different application domains - 1. Module from Ethereum Blockchain, 2. Module from PyTorch deep learning framework, 3. Microsoft SEAL encryption library components and 4. Sed stream editor. We found the classification models for all subject programs resulted in high precision, recall and specificity, averaging to 89\%, 88\% and 92\% respectively, while only training with an average 15\% of the total traces. Our experiments show that the proposed NN model is promising as a test oracle and is able to learn runtime patterns to distinguish test executions for systems and tests from different application domains. 


\end{abstract}

\maketitle

\section{Introduction}
\label{sec:intro}

As the scale and complexity of software increases, the number of tests needed for effective validation becomes extremely large, 
slowing down development, hindering programmer productivity, and ultimately making development costly~\cite{ammann2016introduction, stratis2016test}.
To make testing faster, cheaper and more reliable, it is desirable to automate as much of the process as possible. 

Over the past decades, researchers have
made remarkable progress in automatically generating effective test inputs~\cite{chen2013orchestrated, bertolino2007software}. Automated test input generation tools, however, generate substantially more tests than manual approaches. This becomes an issue when determing the correctness of test executions, a procedure referred to as the \emph{test oracle}, that is still largely manual, relying on developer expertise.  Recent surveys on the test oracle problem~\cite{barr2015oracle, nardi2015survey, langdon2017inferring} show that automated oracles based on formal specifications, metamorphic relations~\cite{liu6613484} and independent program versions are not 
widely applicable and difficult to use in practice. 
In this paper, we seek to address the test oracle problem. More specifically, for a system with a large set of test inputs, that are automatically 
and/or manually generated, we ask,  \\ 
\emph{`` Is there a widely applicable technique that automates the classification of test executions as pass/fail ? ''}
\begin{figure*}[h]
	\centering
	\includegraphics[width = 0.8\textwidth]{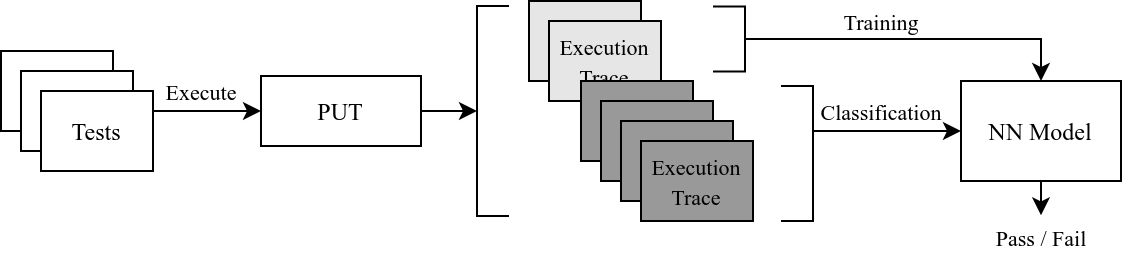}
	\vspace{-6pt}
	\caption{Key idea in our approach.}
	\label{fig:key-idea}
	\vspace{-8pt}
\end{figure*}

\textbf{Key Idea.}
We explore supervised machine learning to infer a test oracle from labelled execution traces of a given system. 
In particular, we use neural networks (NNs), well suited to learning complex functions and classifying patterns, to design the test oracles.
Our technique is widely applicable and easy to use, as it only requires execution traces gathered from running tests through the program under test (PUT) to design the oracle. This is shown in Figure~\ref{fig:key-idea} where a small fraction of the gathered execution traces labelled with pass/fail (shown in light gray) is used to train the NN model which is then used to automatically classify the remaining unseen execution traces (colored dark gray). 

Previous work exploring the use of NNs for test oracles has been in a restricted context -- applied to very small programs with primitive data types, and only considering their inputs and outputs~\cite{vanmali2002using, jin2008artificial}. Information in execution traces which we believe is useful for test oracles has not been considered by existing NN-based approaches.
Other bodies of work in program analysis have used NNs
to predict method or variable names and detecting name-based bug patterns~\cite{alon2018code2vec, pradel2018deepbugs} relying on static program information, namely,  embeddings of the Abstract Syntax Tree (AST) or source code. 
Our proposed approach is the first attempt at using \emph{dynamic execution trace information in NN models for classifying test executions}.

Our approach for inferring a test oracle has the following steps, 
\begin{enumerate}[itemsep = 0pt, topsep = 0pt, partopsep=0pt]
\item Instrument a program to gather execution traces as sequences of method invocations. 
\item Label a small fraction of the traces with their classification decision. 
\item Design a NN component that embeds the execution traces to fixed length vectors.
\item Design a NN component that uses the line-by-line trace information to classify traces as pass or fail.
\item Train a NN model that combines the above components and evaluate it on unseen execution traces for that program. 
\end{enumerate}
The novel contributions in this paper are in Steps 3, 4 and 5. Execution traces from a program vary widely in their length and information. We propose a technique to encode and summarise the information in a trace to a fixed length vector that can be handled by a NN. We then design and train a NN to serve as a test oracle.  

\textbf{Labelled execution traces.} Given a PUT and a test suite, we gather execution traces corresponding to each of the test inputs in the test suite with our instrumentation framework. Effectively learning a NN classifier for a PUT that distinguishes correct from incorrect executions requires labelled data with both passing and failing examples of traces. 
We require a small fraction of the overall execution traces to be labelled, which is likely to be a manual process. 
As a result, our proposed approach for test oracle is \emph{not} fully automated.
We hypothesize that the time invested in labelling a small proportion of the traces is justified with respect to the benefit gained in automatically classifying the remaining majority of traces. 
In contrast, with no classifier, the developer would have had to specify expected output for all the tests, which is clearly more time consuming than the small proportion of tests we need labelled. 



\textbf{NN Architecture.} 
An execution trace in our approach comprises multiple lines, with each line containing information on a method invocation. 
Our architecture for encoding and classifying an execution trace uses multiple components: (1) Value encoder for encoding values within the trace line to a distributed vector representation, (2) Trace encoder encoding trace lines within a variable-length trace to a single vector, and (3) Trace Classifier that accepts the trace representation and classifies the trace. The components in our architecture is made up of LSTMs, one-hot encoders, and a multi-layer perceptron. 

\textbf{Case Studies.} We evaluate our approach using 4 subject programs and tests from different application domains - a single module from Ethereum project~\cite{ethereum}, a module from Pytorch~\cite{pytorch}, one component within Microsoft SEAL encryption library~\cite{sealcrypto} and a Linux stream editor~\cite{sed}.  
One of the 4 subject programs were accompanied by both passing and failing tests that we could directly use in our experiment. The remaining three programs were only accompanied by passing tests. We treated these programs as reference programs. We then generated PUTs by artificially seeding faults into them. We generated traces through the PUTs using the existing tests, labelling the traces as passing or failing based on comparisons with traces from the reference program.  We trained a NN model for each PUT using a fraction of the labelled traces. We found our approach for designing a NN classification model was effective for programs from different domains. We achieved high accuracies in detecting both failing and passing traces, with an average precision of 89\% and recall of 88\%. Only a small fraction of the overall traces (average 15\%) needed to be labelled for training the classification models.

\noindent In summary, the paper makes the following contributions:
\begin{itemize}[itemsep = 0pt, topsep = 0pt, partopsep=0pt]
\item Given a PUT and its test inputs, we provide a framework that instruments the PUT and gathers test execution traces as sequences of method invocations. 
\item A NN component for encoding variable-sized execution traces into fixed length vectors.
\item A NN for classifying the execution traces as pass or fail.  
\item We provide empirical evidence that this approach yields effective test oracles for programs and tests from different application domains. 
\end{itemize}
\vspace{-5pt}

 \section{Background}
 \label{sec:background}
When a test oracle observes a test execution, it returns a
test verdict, which is either pass or fail, depending on whether the observations match 
expected behaviour. A test execution is execution of the PUT with a test input. The importance of oracles as an integral part of the testing process has been a key topic of research for over three decades. 
We distinguish four different kinds of test oracles,
based on the survey by Barr et al.~in 2015~\cite{barr2015oracle}. The most common form of test oracle is a \emph{specified oracle},
one that judges behavioural aspects of the system under test
with respect to formal specifications. Although formal specifications are effective in identifying failures,  
defining and maintaining such specifications is expensive
and also relatively rare in practice.
\emph{Implicit} test oracles require no domain knowledge and 
are easy to obtain at no cost. However, they are
limited in their scope as they are only able to reveal  particular anomalies like buffer overflows, segmentation faults, deadlocks. 
\emph{Derived} test oracles use 
documentations or system executions, to judge a system's behaviour,
when specified test oracles are unavailable. 
However, derived test oracles, like metamorphic relations and inferring invariants, is either not automated  
or it is inaccurate and irrelevant making it challenging to use.

For many systems and much of
testing as currently practised in industry, the
tester does not have the luxury of formal specifications or assertions or even automated partial
oracles~\cite{hierons2009verdict, hierons2012oracles}. 
Statistical analysis and machine learning techniques provide a useful alternative for understanding software behaviour using data 
gathered from a large set of test executions.

\subsection{Machine Learning for Software Testing}
Briand et al.~\cite{briand2008novel}, in $2008$, presented a comprehensive overview of existing techniques that apply machine learning for addressing testing challenges.
Among these, the closest related work is that of
Bowring et al.\ in $2004$~\cite{bowring2004active}. They proposed an active learning approach to
build a classifier of program behaviours using a frequency profile of single events in the
execution trace. 
Evaluation of their approach was conducted over one small program whose specific structure was well suited to their technique. 
Machine learning techniques have also been used in fault detection. 
Brun and Ernst, in $2004$~\cite{brun04finding}, explored the use of support vector machines and decision trees to rank program properties, provided by the user,  that are likely to indicate 
errors in the program. Podgurski et al., in $2003$~\cite{podgurski2003automated},
use clustering
over function call profiles to determine which failure reports
are likely to be manifestations of an underlying error. A training step determines which features are of interest
by evaluating those that enable a model to distinguish
failures from non-failures. The technique does not
consider passing runs. In their experiments, most
clusters contain failures resulting from
a single error.

More recently, Almaghairbe et al.~\cite{almaghairbe2017separating} proposed an unsupervised learning technique to classify unlabelled execution traces of simple programs. They gather two kinds of execution traces, one with only inputs and outputs, and another that includes the sequence of method entry and exit points, with only method names. Arguments and return values are not used. 
They use agglomerative hierarchical clustering algorithms to build an automated test oracle, assuming passing traces are grouped into large, dense clusters and failing traces into many small clusters. They evaluate their technique on 3 programs from the SIR repository~\cite{sir}. The proposed approach has several limitations. They only support programs with strings as inputs. 
They do not consider correct classification of passing traces.
The accuracy achieved by the technique is not high, classifying approximately 60\% of the failures. Additionally, fraction of outputs that need to be examined by the developer is around 40\% of the total tests, which is considerably higher than the labelled data used in our approach. 
We objectively compared the accuracy achieved by the hierarchical clustering technique against our approach using 5 PUTs, discussed in Section~\ref{sec:results}. We found our approach achieves significantly higher accuracy in classifying program executions across all case studies. 
Existing work using execution traces for bug detection has primarily been based on clustering techniques. Neural networks, especially with deep learning, have been very successful for complex classification problems in other domains like natural language processing, speech recognition, computer vision. There is limited work exploring their benefits for software testing problems.

\paragraph{Neural Networks for Test Oracles} 
NNs were first used by Vanmali et al.~\cite{vanmali2002using} in 2002 to simulate behaviour of simple programs
using their previous version, and applied this model to regression testing of unchanged functionalities. 
Aggarwal et al.~\cite{aggarwal2004neural} and Jin et al.~\cite{jin2008artificial} applied the same approach to test a triangle classification program, that computes the relationship among three edge inputs to determine the type of triangle. 
The few existing approaches using NNs have been applied to simple programs having small I/O domains.
The following challenges have not been addressed in existing work, \\ 
 \noindent 1. Training with test execution data and their vector representation -- Existing work only considers program inputs and outputs that are of primitive data types (integers, doubles, characters). 
 Test data for real programs often use complex data structures and data types defined in libraries. There is a need for techniques that encode such data. In addition, existing work has not attempted to use program execution information in NNs to classify tests.  Achieving this will require novel techniques for encoding execution traces and designing a NN that can learn from them.    \\
 \noindent 2. Test oracles for industrial case studies - Realistic programs with complex behaviours and input data structures has not been previously explored. \\
 \noindent 3. Effort for generating labelled training data - Training data in existing work has been over simple programs, like the triangle classification program, where labelling the tests was straightforward. Availability of labelled data that includes failing tests has not been previously discussed. Additionally, the proportion of labelled data needed for training and its effect on model prediction accuracy has not been systematically explored. 

 \paragraph{Deep Learning for Software Testing}
 The performance of neural networks as classifiers was boosted with the birth of deep learning in 2006~\cite{hinton2006fast}.
 Deep learning methods have \emph{not} been explored extensively for software testing, and in particular for the test oracle problem.
 Recently, a few techniques have been proposed for automatic pattern-based bug detection. For example, 
 Pradel et al.~\cite{pradel2018deepbugs} proposed a deep learning-based static analysis for automatic name-based bug detection and
 Allamanis et al.~\cite{allamanis2018learning} used graph-based neural static analysis for detecting variable misuse bugs. 
In addition to these techniques, several other deep learning methods for statically representing code have been developed~\cite{alon2018code2seq,allamanis16convattn}. 
We do not discuss these further since we are interested in execution trace classification and in NNs that use dynamic trace information rather than a static view of the code.

\paragraph{Embedding Execution Traces for Neural Networks}
One of the main contributions in this paper is an approach for embedding information in execution traces as a fixed length vector to be fed into the neural network. There is limited work in using representations of execution traces. Wang et al.~\cite{wang2017dynamic} proposed embeddings of execution traces in 2017. They use execution traces captured as a sequence of variable values at different program points. A program point is when a variable gets updated. Their approach uses recurrent NNs to summarise the information in the execution trace. Embedding of the traces is applied to an existing program repair tool. The work presented by Wang et al. has several limitations - 1. Capturing execution traces as sequences of updates to every variable in the program has an extremely high overhead and will not scale to large programs. The paper does not describe how the execution traces are captured, they simply assume they have them. 2. The approach does not discuss how variables of complex data types such as structs, arrays, pointers, objects are encoded. It is not clear if the traces only capture updates to user-defined variables, or if system variables are also taken into account. 
3. The evaluation uses three simple, small  programs (eg. counting parentheses in a string) from students in an introductory programming course. 
The complexity and scale of real programs is not assessed in their experiments. Their technique for capturing and directly embedding traces as sequences of updates to every variable is infeasible in real programs. 
Our approach captures and embeds traces as sequences of method invocations and updates to global variables, which scales better than tracking every program variable. We have implemented our instrumentation in the LLVM compiler framework that is language agnostic and scales to industry-sized programs. We support all types of variables and objects, including system defined variables. 


\section{Approach}
 \label{sec:approach}
\input{Approach-arX}

\section{Experiment}
 \label{sec:experiment}
In our experiment, we evaluate the feasibility and accuracy of the NN architecture proposed in Section~\ref{sec:approach} to classify execution traces for 4 subject programs and their associated test suites.
We investigate the following questions regarding feasibility and effectiveness:

\textbf{Q1. Precision, Recall and Specificity:} 
\textit{What is the precision, recall and specificity achieved over the subject programs? }

To answer this question, we use our tool to instrument the source code to 
record execution traces as sequences of method invocations, arguments and return values. 
A small fraction of the  execution traces are labelled (\emph{training set}) and fed to our framework to infer a classification model. We then evaluate precision, recall and specificity achieved by the model over unseen execution traces (\emph{test set}) for that program. The test set includes both passing and failing test executions. We use \emph{Monte Carlo cross-validation}, creating random splits of the dataset into training and test data. We created 15 such random splits and averaged precision, recall and specificity computed over them. In our experiments, we do not use a validation set to tune the hyper-parameters in the NN model.  

\textbf{Q2. Size of training set:}
\textit{How does size of the training set affect precision and recall of the classification model?}

For each program, we vary the size of training set from 5\% to 30\% of the overall execution traces and observe its effect on precision and recall achieved. 

\textbf{Q3. Comparison against state of art:}
\textit{How does the precision, recall and specificity achieved by our technique compare against agglomerative hierarchical clustering, proposed by Almaghairbe et al.~\cite{almaghairbe2017separating} in 2017? }

We choose to compare against the hierarchical clustering work as it is the most relevant and recent in classifying execution traces. Traces used in their work are sequences of method invocations, similar to our approach. Other test oracle work that use NNs is not used in the comparison as they do not work over execution traces, and are limited in their applicability to programs with numerical input and output which is not the case for programs in our experiment.

All experiments are performed on a single machine with 4 Intel i5-6500 CPU cores, Nvidia RTX 2060 GPU, 16GB of memory. 

\subsection{Labelling Traces}
\label{sec:labelling-traces}
All our subject programs are open source, and most of them were only accompanied by passing tests. This is not uncommmon as most released versions of programs are correct for the given tests. We take these correct programs to be reference implementations. To enable evaluation of our approach that distinguishes correct versus incorrect executions, we need subject programs with bugs. We, therefore, generate PUTs by automatically mutating the reference implementation using common mutation operators~\cite{jia2011analysis} listed below, 
\begin{enumerate}[itemsep = 0pt, topsep = 0pt, partopsep=0pt]
	\item {Arithmetic operator replacement applied to \{$+, -, *, /,\\ --, ++$\}.}
	\item {Logical connector replacement applied to \{$\&\&, ||, !$\}.}
	\item {Bitwise operator replacement applied to \{$\&, |, \wedge, ~, <<, >>$\}.}	\item {Assignment operator replacement applied to \\ \{$+=, -=, *=, /=, \%=, <<=, >>=, \&=, |=, \wedge=$\}.}
\end{enumerate}
\begin{figure}[ht!]
	\vspace{-12pt}	
	\centering
	\includegraphics[width = 0.6\columnwidth]{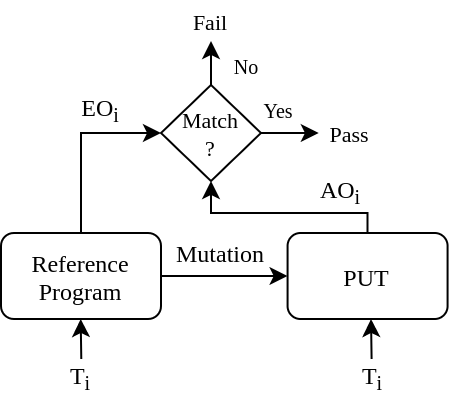}
	\caption{Labelling test executions by matching actual and expected behavior.}
	\label{fig:labelling}
	\vspace{-10pt}
\end{figure}
A PUT is generated by seeding a single fault into the reference implementation at a random location using one of the above mutation operators. 
We used an independent open source mutation tool\footnote{\url{https://github.com/chao-peng/mutec}} to generate PUTs from a given reference program. 
Figure~\ref{fig:labelling} shows a PUT generated by seeding a single fault into a reference program.
As seen in Figure~\ref{fig:labelling}, we run each test, $T_i$, in the test suite, through both the reference program and PUT, and label the trace as \emph{passing} if the expected output, $EO_i$, from the reference matches the actual output, $AO_i$, from the PUT. If they do not match, the trace is labelled as \emph{failing}.  
We rejected PUTs from mutations that did not result in any failing traces (outputs always match with the reference). This avoids the problem of equivalent mutants. 
All the PUTs in our experiment had both passing and failing traces. 

\subsection{Subject Programs}
\label{sec:subj-programs}
We chose subject programs from different domains to assess applicability of our approach, namely from the blockchain, deep learning, encryption and text editing domains. A description of the programs and associated tests is as follows. \\
\noindent\textbf{1. Ethereum}~\cite{ethereum} is an open-source platform based on blockchain technology, which supports smart contracts. Within it, we evaluate our approach over the \texttt{Difficulty} module that calculates the mining difficulty of a block, in relation to different versions (eras) of the cryptocurrency (Byzantium, Homestead, Constantinople etc.). The calculation is based on five fields of an \texttt{Ethereum} block, specified in the test input. 

\paragraph{Tests}

We use the default test inputs provided by Ethereum's master test suite for the \texttt{Difficulty} module. We test this module for the Byzantium era of the cryptocurrency (version 3.0). The test suite contains 2254 \emph{automatically} generated test inputs. Each test input contains one hex field for the test input of the difficulty formula and another hex field for the expected output of the program. All the test inputs provided with the module are passing tests with the actual output equal to the expected output. As a result, we use the provided module as a reference implementation. As described in Section~\ref{sec:labelling-traces}, we seed faults into the reference implementation to generate PUTs, each containing a single mutation. For the difficulty module, we generate 2 PUTs -- 1. Ethereum-SE with a seeded fault in the core functionality of the difficulty module, and 2. Ethereum-CD  with a fault seeded in one of the functions that is external to the core function but appears in the call graph of the module.  The balance between passing and failing tests varies between the two PUTs, Ethereum-CD being perfectly balanced and Ethereum-SE being slightly imbalanced (828 failing and 1426 passing tests). 

\noindent\textbf{2. Pytorch}~\cite{pytorch} is an optimized tensor library for deep learning, widely used in research. In our experiment, we evaluate our model over the \texttt{intrusive\_ptr} class, which implements a pointer type with an embedded reference count.
We chose this class because it had a sizeable number of tests (other modules had $<20$ published tests). 

\paragraph{Tests}
Implementation of the class is accompanied by 638 tests, all of which are passing. We, thus, use this as the reference implementation.  
As with \texttt{Ethereum}, we apply mutations to the \texttt{intrusive\_ptr} implementation to generate a single PUT. Upon comparison with the reference,  318 of the existing tests are labelled passing through the PUT and 320 as failing.  

\noindent\textbf{3. Microsoft SEAL}~\cite{sealcrypto} is an open-source encryption library. In our experiment, we study one component within Microsoft SEAL, the \texttt{Encryptor} module, which is accompanied by tests. This component is responsible for performing data encryption.


\paragraph{Tests}
The \texttt{Encryptor} component is accompanied by 133 tests. 
The provided tests were all passing tests, with matching expected and actual output. 
As with previous programs, we generate a PUT by mutating the original implementation. On the PUT, 11 tests fail and 122 pass. 


\noindent\textbf{4. Sed}~\cite{sed} is a Linux stream editor that performs text transformations on an input stream.
\paragraph{Tests}
We use the fifth version of \texttt{Sed} available in the SIR repository~\cite{sir}. This version is accompanied by 370 tests, of which 352 are passing and 18 are failing. The failing tests  point to real faults in this version. Since the implementation was accompanied by both passing and failing, we used it as the PUT. We did \emph{not} seed faults to generate the PUT. 

\paragraph{Checks to avoid data leakage}
We ensure no test oracle data is leaked into traces. We remove expected outputs, assertions, exceptions, test names and any other information that may act directly or indirectly as a test oracle. For example, Ethereum uses \texttt{BOOST} testing framework to deploy its unit tests. We remove expected outputs and assertions in the test code that compare actual with the expected output e.g. \texttt{BOOST\_CHECK\_EQUAL}. 

For PUTs generated by seeding faults into the reference implementation, we only use one PUT for each reference implementation except in the case of Ethereum where we generated two PUTs, since faults were seeded in different files. Generating more PUTs for each reference implementation would be easy to do. However, we found our results across PUTs for a given reference program only varied slightly. As a result, we only report results over one to two PUTs for each reference implementation. 

\subsection{Performance Measurement}
For each PUT, we evaluate performance of the classification model over unseen execution traces. As mentioned in Section~\ref{sec:NN-model}, we use positive labels for failing traces and negative labels for passing. We measure 
\begin{enumerate}
 \item \emph{Precision} as the ratio of number of traces correctly classified as ``fail'' (\texttt{TP}) 
to the total number of traces labelled as ``fail'' by the model (\texttt{TP + FP}). 
 \item \emph{Recall} as the ratio of failing traces that were correctly identified  (\texttt{TP/(TP + FN)}). 
 \item \emph{Specificity} or true negative rate (TNR) as the ratio of passing traces that were correctly identified (\texttt{TN /(TN + FP)}).
\end{enumerate}
\texttt{TP, FP, TN, FN } represent true positive, false positive, true negative and false negative, respectively.  

\subsection{Hierarchical Clustering}
In research question 3 in our experiment, we compare the classification accuracy of our approach against agglomerative hierarchical clustering proposed by Almaghairbe et al.~\cite{almaghairbe2017separating}. Their technique also considers execution traces as sequences of method calls, but only encoding callee names. Caller names, return values and arguments are discarded. We attempted to add the discarded information, but found the technique was unable to scale to large number of traces due to both memory limitations and a time complexity of $\mathcal{O}(n^3)$ where \texttt{n} is the number of traces. For setting clustering parameters for each subject program, we evaluate different types of linkage (\texttt{single}, \texttt{average}, \texttt{complete}) and a range of different cluster counts (as a percentage of the total number of tests): 1, 5, 10, 20 and 25\%. We use Euclidean distance as the distance measure for clustering. For each program, we report the best clustering results achieved over all parameter settings. 

 \section{Results and Analysis}
 \label{sec:results}

In this section, we present and discuss our results in the context of the research questions presented in Section~\ref{sec:experiment}.

\begin{table*}[]
	\centering
	\small
	\begin{tabular}{|l|l|l|l|l|l|l@{\hskip 5mm}|l|l|l|}
		\hline
		PUT & Lines of & \% Traces & Total & \multicolumn{3}{c|}{Our Approach} & \multicolumn{3}{c|}{Hierarchical Clustering~\cite{almaghairbe2017separating}}\\
		& Code & for training &  \# Traces      & {Precision} & {Recall} & {TNR} & {Precision} & {Recall} & {TNR}\\ 
		\hline
		Ethereum-CD & 55927 & 15 & 2254 & 0.80 & 0.82 & 0.79 & 1.0 & 0.49 & 1.0 \\ 
		Ethereum-SE & 55927 & 15 & 2254 & 0.99 & 0.82 & 0.86 & 1.0 & 0.25 & 1.0 \\ 
		Pytorch & 21090 & 10 & 638 & 0.99 & 0.98 & 0.99 & 0.48 & 1.0 & 0.16 \\
		SEAL Encryptor & 25967 & 30 & 132 & 0.75 & 0.86 & 0.98 & 0.16 & 0.36 & 0.83 \\ 
		Sed & 4492 & 10  & 370 & 0.94 & 0.94 & 0.99 & 0.35 & 0.63 & 0.86 \\ \hline
	\end{tabular}
	\caption{Precision, Recall and True Negative rate (TNR) using our approach and hierarchical clustering.} 
	\vspace{-17pt}
	\label{tab:results}
\end{table*}

\subsection{Q1. Precision, Recall and Specificity} 

Table~\ref{tab:results} shows the precision, recall and specificity achieved by the classification models in our approach for the different PUTs. Results with the hierarchical clustering approach by Almaghairbe et al.~\cite{almaghairbe2017separating} are also presented in Table~\ref{tab:results} for comparison,  but this is discussed in Q3 in Section~\ref{sec:q3}. 
The column showing \% of traces used in training varies across programs, we show the lowest percentage that is needed to achieve {near maximal precision and recall}.  

The classification models for all 5 PUTs achieve more than $75\%$ precision and recall, with an average of $89\%$ and $88\%$, respectively. Our technique works particularly well for Pytorch and Sed, achieving $>=94\%$. This implies that the number of false positives in the classification is very low and a large majority of the failing traces are correctly identified. 

The classification models for all PUTs also achieve high specificity ($> = 79\%$, average $92\%$). This implies that the NN models are able to learn runtime patterns that distinguish not only failing executions, but also passing executions with a high degree of accuracy. These results are unprecedented as we are not aware of any technique in the literature that can classify both passing and failing executions at this level of accuracy.

\paragraph{Analysis}
To understand the results in Table~\ref{tab:results}, for each of the PUTs, we inspected and compared passing and failing traces using a combination of longest common subsequence, syntactic diffs, and manual inspection. We also performed \emph{ablation} - systematically removing information (one parameter at a time) from the traces, training new classification models with the modified traces and observing their effect on precision, recall and specificity (TNR). In our experiments, we systematically remove the following parameters from the original traces -- function call names, arguments, and return values. Table~\ref{tab:sec_removed} shows the results from our ablation study.
We discuss results for each of the programs in the following paragraphs. 

Over SEAL Encryptor, our approach achieves 75\% precision, 86\% recall and 98\% specificity when trained with 30\% of the traces. Encryptor requires a higher fraction of traces for training when compared to other PUTs, as the number of failing traces is very small ($= 11$), unlike other programs. Although we handle imbalance in datasets by weighting samples in the loss function, the NN still needs some representatives of the failing class during training. Using 10\% of the traces in training, will only provide one example of failing trace (10\% of 11)  which is not enough for the NN model to learn to distinguish failing versus passing behaviour. Training using 30\% of the traces includes 3 failing traces which allows the NN to achieve 75\% precision. High precision with only 3 failing traces is because all the failing traces for this program have the same call sequence, which is sufficiently different from passing traces. Passing traces do not all have the same sequence. However, due to the availability of a larger set of passing traces (training with 30\% is 40 passing traces), the NN is able to identify the different method call patterns in passing traces accurately (98\% specificity).  The ablation study in Table~\ref{tab:sec_removed} shows that all the parameters contribute to model performance as removing them has a detrimental effect. 

For PyTorch, we achieve 99\% precision, 98\% recall and 99\% specificity when trained with 10\% of the traces. The dataset for PyTorch PUT is balanced (318 passing and 320 failing). 10\% of the traces during training provides sufficient examples from both passing and failing classes for the NN to learn to distinguish them. We find the reason for the superior performance of our model over PyTorch is because all failing traces have significantly fewer trace lines than passing traces. The consistent difference in length of traces between the two classes allows the NN to easily distinguish them. The ablation study in Table~\ref{tab:sec_removed} shows arguments in traces matter for model performance, while  method names and return values are irrelevant.  

With Sed, our model achieves 94\% precision and recall, and 99\% specificity using 10\% of the traces in training. The dataset for Sed is unbalanced, with only 18 failing and 352 passing. 10\% of the traces in training uses 2 failing tests and 35 passing tests. Given the extremely small sample of failing tests, it is surprising that the model classifies and identifies failing traces with such high precision and recall. To understand this, we examined both the passing and failing trace lines. We find the length of passing and failing traces is similar. All failing traces, however, have a call to a function, \texttt{getChar}, towards the end of the trace. This function call is absent in passing traces. We believe associating this function call to failing traces may have helped the performance of the NN. The ablation study in Table~\ref{tab:sec_removed} shows all the parameters considered in our traces are important for model performance. 

For Ethereum-CD, our model achieves 80\% precision, 82\% recall and 79\% specificity when trained with 15\% of the traces - 169 passing and 169 failing. Ethereum-CD was generated from the reference implementation using an arithmetic operator mutation in a function deeply embedded in the call graph for the difficulty module. Differences between failing and passing traces are not apparent, and analysing longest common subsequence, syntactic diff and manual inspections did not reveal any characteristic feature for failing or passing traces. We believe the model performance of around 80\% precision, recall and specificity is due to the similarity between passing and failing traces and the esoteric nature of the mutation. Ablation study for this program reveals that all features in the traces slightly impact model performance. 

For Ethereum-SE, our model achieves 99\% precision, 82\% recall and 86\% specificity with 15\% traces in training - 214 failing and 124 passing. 
Unlike Ethereum-CD, mutation to generate Ethereum-SE was in the core functionality. Failing traces when compared to passing traces had differences towards the end of the trace which is easily distinguished by the NN. Curiously, removing return values in the ablation study, increases recall and specificity. This may be because the model was previously overfitting to return values in traces which may not have been relevant to the classification. 

\paragraph{Summary}
Overall, we find NN models for all our PUTs perform well as a test oracle, achieving an average of 89\% precision, 88\% recall and 92\% specificity. The NN models perform exceptionally well for programs whose traces have characteristic distinguishing features between passing and failing executions, such as differences in trace lengths or presence of certain function call patterns. In the absence of such features, NNs can still do well if it has enough training samples, as in Ethereum-CD. We also find our approach can cope effectively with unbalanced datasets -- three of the five programs in our experiment have unbalanced passing and failing traces. 

\begin{table}[!h]
	\centering
	\footnotesize
	\begin{tabular}{|p{1.5cm}|l|c|c|c|}
		\hline
		PUT & Omitted Info. & P & R & TNR \\ \hline
		\multirow[t]{4}{*}[-0.05cm]{\textbf{Ethereum-CD}} 
		& Function names & 0.63 & 0.64 & 0.62 \\
		& Return values & 0.68 & 0.87 & 0.60 \\ 
		& Arguments & 0.54 & 0.78 & 0.35 \\ 
		\hline

		\multirow[t]{4}{*}[-0.05cm]{\textbf{Ethereum-SE}} 
		& Function names & 0.96 & 0.84 & 0.35 \\
		& Return values & 0.99 & 0.97 & 0.93 \\ 
		& Arguments & 0.96 & 0.84 & 0.33 \\ 
		\hline
		
		\multirow[t]{4}{*}[-0.05cm]{\textbf{Pytorch}} 
		& Function names & 0.99 & 1.0 & 1.0 \\
		& Return values & 0.99 & 0.99 & 0.99 \\ 
		& Arguments & 0.51 & 0.99 & 0.04 \\ 
		\hline
		
		\multirow[t]{4}{*}[-0.315cm]{\textbf{\shortstack[l]{Seal\\Encryptor}} } 
		& Function names & 0.53 & 0.87 & 0.92 \\
		& Return values & 0.46 & 0.99 & 0.90 \\ 
		& Arguments & 0.28 & 0.88 & 0.76 \\ 
		\hline

		\multirow[t]{4}{*}[-0.05cm]{\textbf{Sed}} 
		& Function names & 0.19 & 0.72 & 0.24 \\
		& Return values & 0.48 & 0.52 & 0.85 \\ 
		& Arguments & 0.30 & 0.40 & 0.73 \\ 
		\hline

	\end{tabular}
	\caption{Precision (P), Recall (R) and Specificity (TNR) for each PUT omitting certain trace information.}
	\vspace{-5pt}
	\label{tab:sec_removed}
\end{table}

\subsection{Q2. Size of training set}
\label{sec:q2}
\begin{figure*}[!htbp]

	\centering
	\setlength{\tabcolsep}{1em}
	\setlength{\extrarowheight}{20pt}
	\begin{tabular}{ccc}
        \includegraphics[scale=0.18]{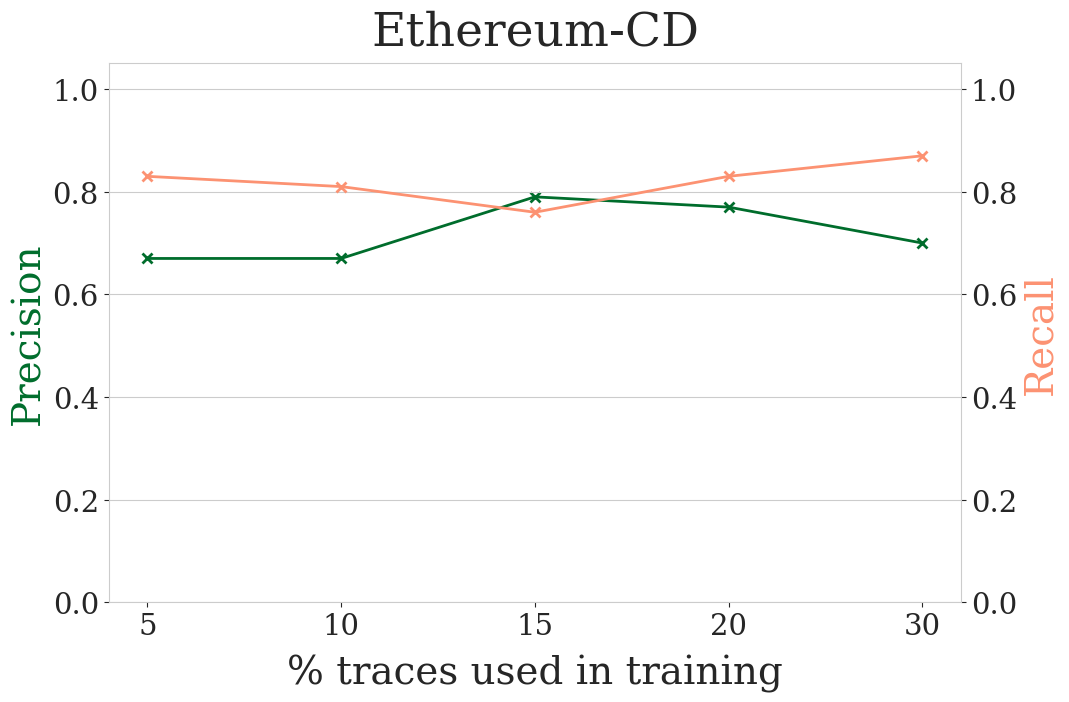}
        &
        \includegraphics[scale=0.18]{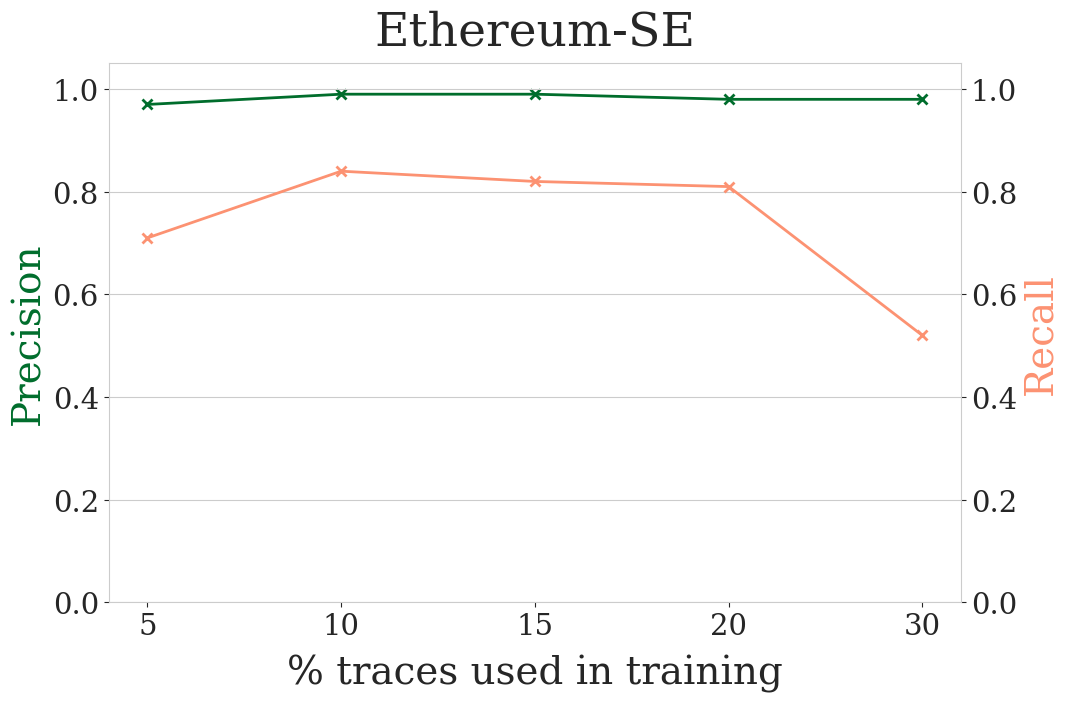}
        &
        \includegraphics[scale=0.18]{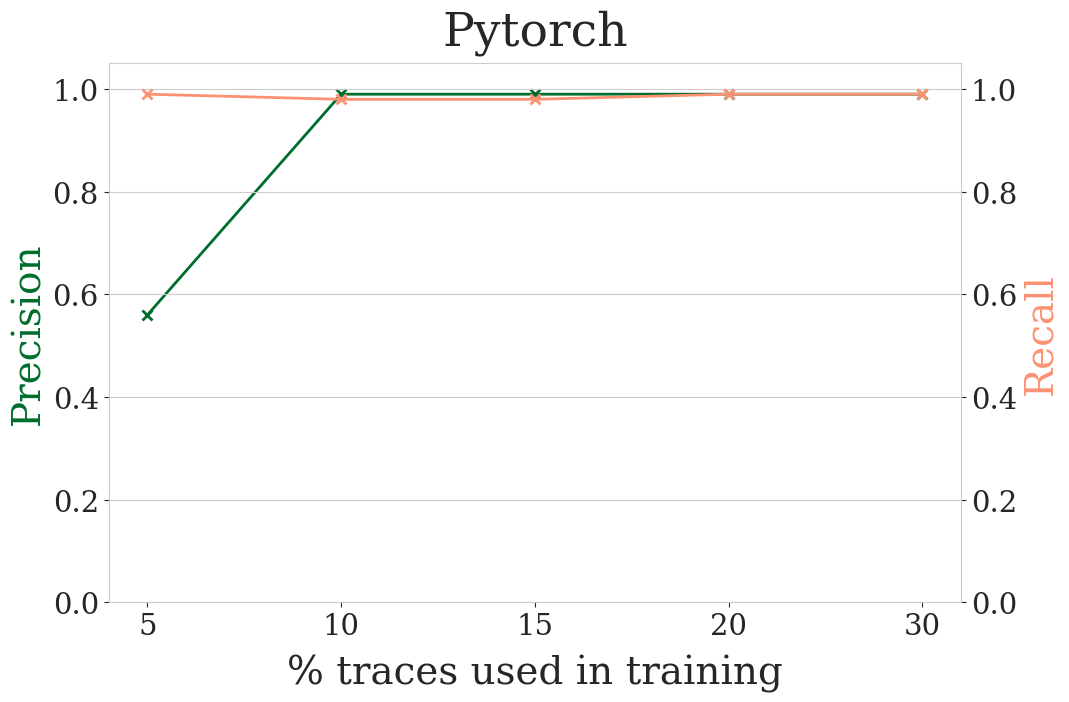}
        \\ [0.1cm]
		\includegraphics[scale=0.18]{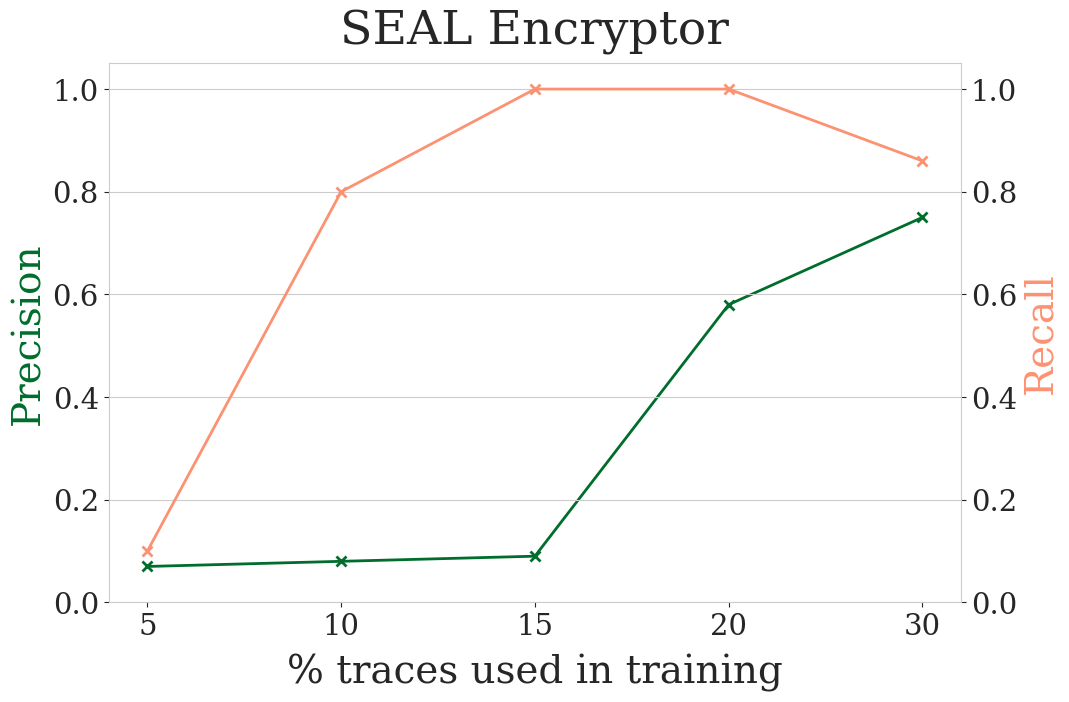}
		&
		\includegraphics[scale=0.18]{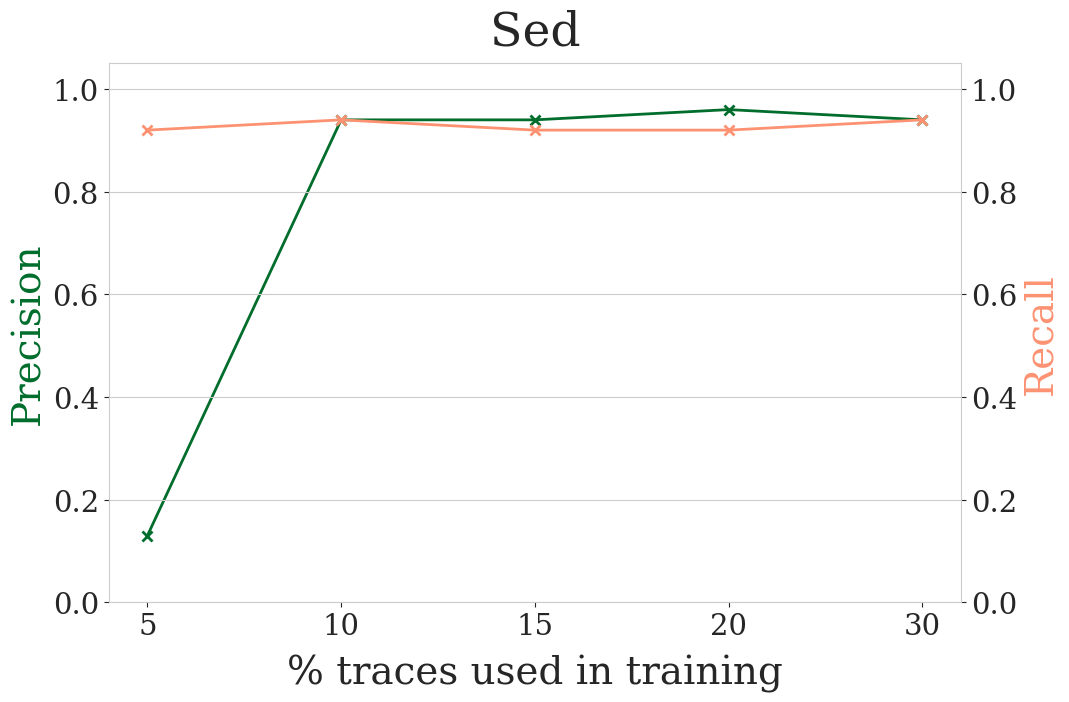}
		\\[0.1cm]
	\end{tabular}  
	\vspace{-10pt}
	\caption{Precision and recall achieved by classification model over each PUT.}
	\vspace{-5pt}
	\label{fig:precision-recall}

\end{figure*}


Figure~\ref{fig:precision-recall} shows precision and recall achieved by our approach with different training set sizes.
The fraction of traces needed in training to achieve near maximal performance was 10\% to 30\% across the PUTs.
Excluding SEAL Encryptor, all the other programs only needed to be trained over 15\% of the traces to achieve near maximal performance. SEAL encryptor had very few failing traces, requiring a larger fraction of traces to get sufficient representation of failing classes during training. 
As seen in the plots in Figure~\ref{fig:precision-recall}, increasing the \% of traces used in training does not increase precision and recall for all PUTs. For instance, Pytorch and Sed observe a dramatic increase in precision and recall  when going from 5 to 10\% traces in training. Performance, however, stagnates after that point with increasing traces. With Ethereum-CD and Ethereum-SE, precision or recall becomes worse after 20\% traces. This maybe because the model is overfitting to the training traces.  

It is also worth noting that the absolute size of our training set varies across subject programs. We find our approach works with training sets with as few as 3 failing traces to as many as 214. The range of passing tests in training was between 31 and 169.
\subsection{Q3. Comparison against state of art}
\label{sec:q3}

Table~\ref{tab:results} presents precision, recall, and specificity (TNR) achieved by the agglomerative hierarchical clustering proposed by Almaghairbe et al.~\cite{almaghairbe2017separating} on each of the PUTs. Comparing the precision, recall and TNR of our approach versus hierarchical clustering, we find our approach clearly outperforms the clustering approach on all but the Ethereum-CD PUT. This is because the hierarchical clustering assumption does not hold for these programs. According to this assumption, passing traces tend to be grouped in a few big clusters and failing traces are grouped into many small clusters. However, for these programs, passing traces tend to be grouped in many small clusters based on their call sequence pattern, making it hard to distinguish them from failing traces by simply comparing cluster sizes.

\begin{figure}[ht!]
\vspace{-8pt}
\centering
\includegraphics[scale=0.24]{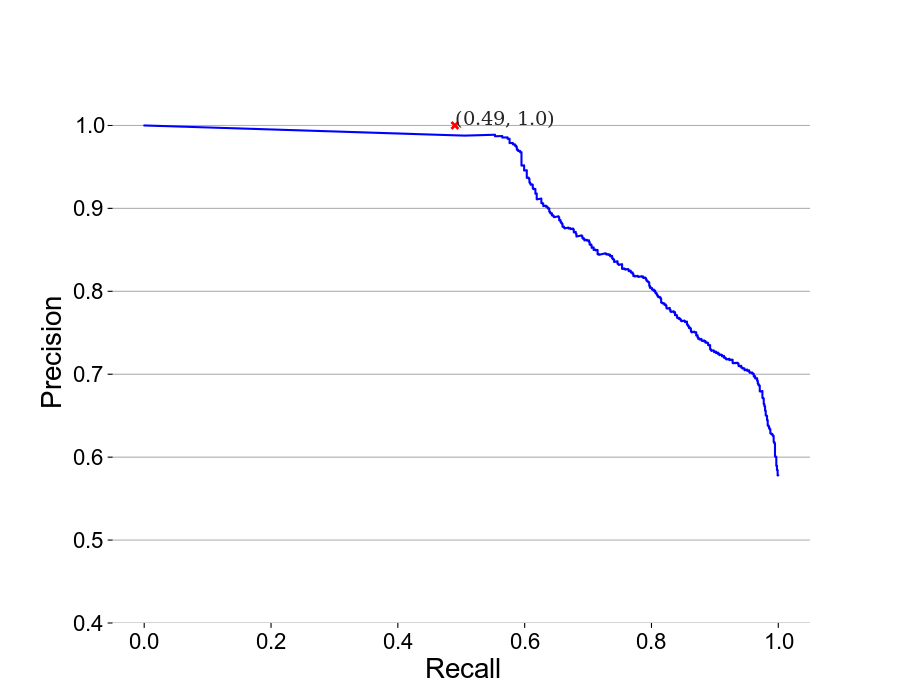}
\vspace{-8pt}
\caption{Precision-Recall curve for \texttt{Ethereum-CD}.}
\label{fig:roc}
\vspace{-5pt}
\end{figure}

With Ethereum-CD, the hierarchical clustering approach achieves precision and specificity of 100\% and a recall of 49\%. This is achieved with complete-linkage clustering, Euclidean distance and a cluster count equal to 10\% of total traces. In contrast our approach achieves a precision of 80\%, recall of 82\% and specificity of 79\%. To enable better comparison, we plot the precision-recall curve of the NN model in Figure~\ref{fig:roc} for Ethereum-CD, using 15\% of the traces in training. 

This curve shows the precision and recall of our trained model with respect to different values of the classification threshold.  It is clear from the plot that for the same value of recall (49\%), hierarchical clutering performs marginally better than our approach - 100\% versus 99\%. Hierarchical clustering works well over the Ethereum-CD PUT because the traces are clustered into just one big passing cluster and one failing cluster. Lack of cluster fragmentation improved accuracy of the hierarchical clustering approach. Nevertheless, our model achieves comparable performance for such traces. In addition, our model allows trade off between precision and recall by changing the classification threshold which may be driven by requirements or priorities of the use case. This tradeoff is not possible with the clustering approach.

\subsection{Threats to Validity}
\label{sec:threats}
We see three threats to validity of our
experiment based on the selection of subject programs and associated tests. 

First, PUTs for 3 out of the 4 subject programs in our experiment were generated by seeding faults into a reference implementation. A reference implementation with only passing tests is not suitable for evaluating our approach.  To address this, we generated a faulty implementation and ran the original tests through the PUT to gather both passing and failing traces.  
It is possible using real faults in place of seeded faults may lead to
different results. However, Andrews et al. have shown the use of seeded faults leads to conclusions
similar to those obtained using real faults~\cite{andrews2006using, Do1707670}. 
For one of the subject programs, Sed,  we did not artificially seed faults, but instead used the existing implementation as it was accompanied by both passing and failing tests. 

Second, the number of tests that accompanied our subject programs was not very large, ranging from 132 to 2254 tests. The NN models in our experiments produced good performance with small to medium sized test suites that may be automatically or manually generated. Our approach is constrained by the amount of training data and not by the size of the test suite. As a result for programs accompanied by large test suites, the NN model will need a larger training set (fraction of traces to be used in training might still be 15\%). Nevertheless, the labelling effort for a fraction of the tests in our approach is still less than the current practise of labelling all the tests. 


Finally, we conducted our study on subject programs from 4 different application domains which is not representative of all application domains. 
Given that our approach has no domain specific constraints, we believe it will be widely applicable. 

\section{Conclusion}
\label{sec:future}
In this paper, we propose a novel approach for designing a test oracle as a NN model, learning from execution traces of a given program. 
We have implemented an end to end framework for automating the steps in our approach, (1) Gathering execution traces as sequences of method invocations, (2) Encoding variable length execution traces into a fixed length vector, (3) Designing a NN model that uses the trace information to classify the trace as pass or fail. 

We evaluated the approach using 5 realistic PUTs and tests. We found the classification model for each PUT was effective in classifying passing and failing executions, achieving an average of 89\% precision, 88\% recall and 92\% specificity while only training with an average 15\% of the total traces. 
We outperform the hierarchical clustering technique proposed in recent literature by a large margin of accuracy for 4 out of the 5 PUTs, and achieved comparable performance for the other PUT.

\paragraph{Practical use} Our approach can be applied out of the box for classifying tests for any software that can be compiled to LLVM IR. We gather execution traces for test inputs automatically, and require a small fraction of the traces to be labelled with their pass or fail outcomes (average 15\% in our experiments). The remaining traces will then be classified automatically. Our approach is promising with high accuracy and has clear benefits over current industry practices where developers label \emph{all} the tests. Our future work will focus on methods to improve the classification accuracy while reducing the training data requirement using techniques like transfer learning. 

\bibliographystyle{plain}
\bibliography{References}

\end{document}